\RequirePackage{ifpdf}
\pdfoutput=1 
\documentclass{JINST}

\title{Vertex-Detector R\&D for {CLIC}}

\author{D. Dannheim$^a$\thanks{Talk presented at the 13th Topical Seminar on Innovative Particle and Radiation Detectors (IPRD13), 7 - 10 October 2013,  Siena, Italy.}\\
On behalf of the CLIC detector and physics study (CLICdp)
\\
\llap{$^a$}CERN,
  Geneva, Switzerland\\
E-mail: \email{dominik.dannheim@cern.ch}}

\abstract{
A detector concept based on  hybrid planar pixel-detector technology is under development for the CLIC vertex
detector. It comprises fast, low-power and small-pitch readout ASICs implemented in 65 nm CMOS technology (CLICpix) coupled to ultra-thin sensors via low-mass interconnects. The power dissipation of the readout chips is reduced by means of power pulsing, allowing for a cooling system based on forced gas flow. In this paper 
the CLIC vertex-detector requirements are reviewed
and the current status of R\&D on sensors, readout and detector integration is presented.
}


\begin{document}

\section{Introduction}\label{sec:intro}
The proposed Compact LInear Collider (CLIC) concept of a linear
electron-positron collider with a centre-of-mass energy of up to 3~TeV
has a large physics potential, complementing and extending the
measurements of the current LHC experiments~\cite{CDRVol1, CDRVol2, CDRVol3}. It will allow for
precision measurements of standard model physics (e.g. Higgs, top) and
of new physics potentially discovered at the LHC (e.g. SUSY).
Moreover, direct and indirect searches for new physics over a large
range of mass scales will be performed. The demands for precision
physics in combination with the challenging experimental conditions at
CLIC have inspired a broad detector R\&D program. In particular the
vertex-detector systems have to fulfil unprecedented requirements in
terms of material budget and spatial resolution in a location
close to the interaction point, where the rates of beam-induced
background particles are very high. The ongoing
CLIC vertex-detector studies focus on ultra-thin hybrid pixel
detectors and aim for integrated solutions taking into account
constraints from mechanics, power delivery and cooling.

\section{The CLIC machine environment}\label{sec:clic-machine}
The CLIC project studies the feasibility of a linear electron-positron
collider optimized for a centre-of-mass energy of 3 TeV with an
instantaneous luminosity of a few times $10^{34}$cm$^{-2}$s$^{-1}$,
using a novel technique
called two-beam acceleration~\cite{CDRVol1}.
A drive beam of rather low energy but high
current is decelerated, and its energy is transferred to the
low-current main beam, which gets accelerated with gradients of
100~MV/m. The two-beam acceleration scheme thus removes the need for
RF power sources along the accelerating main LINAC.
It is expected that the machine will be
built in several stages with centre-of-mass energies ranging from
a few hundred GeV up to the maximum of 3~TeV, corresponding to an overall length of
the accelerator complex up to 48~km.
In order
to reach its design luminosity of $6\times10^{34}$cm$^{-2}$s$^{-1}$
at a maximum
centre-of-mass energy of 3~TeV, CLIC will operate with very small
bunch sizes
($\sigma_x \times \sigma_y \times \sigma_z \approx 40 \mathrm{nm}
\times 1 \mathrm{nm} \times 44 \mathrm{\mu m}$).
Accelerating structures of 12~GHz drive the two main beams and collisions
occur in bunch crossings (BX) every 0.5~ns for a train duration of 156~ns. The train
repetition rate is 50~Hz.
The short train duration implies that
triggerless readout of the detectors, once
per train, will be implemented.  The power consumption of the detectors,
and therefore the material required for cooling infrastructure, can be
reduced by switching off parts of the frontend electronics during the
20~ms gaps between trains.

\section{Beam-induced backgrounds}\label{sec:backgrounds}
The very small beam sizes lead to strong electromagnetic radiation
(Beamstrahlung) from the electron and positron bunches in the field of
the opposite beam.
The creation of the Beamstrahlung photons reduces
the available centre-of-mass energy of the
$e^+e^-$ collisions and their interaction
leads to lepton pairs and hadrons, most of which are produced
at very low polar angles and are therefore contained in the beam-pipe 
by the axial magnetic field~\cite{background-note}.
The dominant backgrounds in the inner
detectors are incoherently produced electron-positron pairs
(approximately 60 particles / BX) and $\gamma\gamma
\rightarrow$hadrons events (approximately 54 particles / BX).  The
electron-positron pairs are predominantly produced at very small
transverse momenta and low polar angles. The detector occupancies in
the innermost layers can therefore be reduced to an acceptable level
by a careful design optimisation of the inner- and forward-detector
regions. The central beam-pipe walls have to be placed outside the high-rate
region and the inner detectors have to be efficiently shielded from
back-scattered particles originating from the forward region.
The particles produced in $\gamma\gamma \rightarrow$hadrons
interactions, on the other hand, show a harder transverse momentum
spectrum and a more central polar-angle distribution, resulting in
large rates of background particles with every bunch crossing,
reaching the outer detector layers.
While at most one interesting physics event is expected in each train,
$\approx 1000$ hadronic background events are produced.
Pile-up rejection algorithms
based on hit time stamping are needed to separate the physics from the
background events.

The radiation exposure of the main detector elements is expected to be
small, compared to the corresponding regions in high-energy
hadron-colliders. For the non-ionizing energy loss (NIEL), a maximum
total fluence of less than $10^{11}$n$_{eq} / $cm$^2$/year
 is expected for the
inner barrel and forward vertex layers. The simulation results for the
total ionizing dose (TID) predict approximately 200~Gy/year for the
vertex-detector region.

\section{Vertex-detector requirements}\label{sec:vtx-requirements}
The primary purpose of the CLIC vertex detector is to
allow for efficient tagging of heavy quarks through a precise
determination of displaced vertices. Monte Carlo simulations show that
these goals can be met with a high-momentum term in the transverse
impact-parameter resolution of $a\approx 5\mathrm{\mu m}$ and a
multiple-scattering
term of $b\approx 15\mathrm{\mu m}$, using the canonical parametrization
\begin{equation}
\sigma(d_0)=\sqrt{a^2+b^2\cdot\mathrm{GeV}^2/(p^2 \mathrm{sin}^3\theta)},
\end{equation}
where $p$ is the momentum of the particle and $\theta$ is the polar
angle with respect to the beam axis. These requirements on the
measurement precision exceed the results achieved in any of the
currently existing full-coverage vertex systems.  They can be met with
multi-layer barrel and endcap pixel detectors operating in a magnetic
field of 4-5~T and using sensors with a
single-point resolution of $\approx 3\mathrm{\mu m}$ and a material
budget at the level of $<0.2\%$ of a radiation length ($\mathrm{X_0}$)
for the beam-pipe and for each of the detection layers. The
single-point resolution target can be met with pixels of $\approx 25
\mathrm{\mu m} \times 25 \mathrm{\mu m}$ and analog readout.  The
material-budget target corresponds to a thickness equivalent to less
than 200~$\mathrm{\mu}$m of silicon, shared by the active material, the
readout, the support and the cooling infrastructure. This implies that
no active cooling elements can be placed inside the vertex
detector. Instead, cooling through forced air flow is foreseen,
limiting the maximum power dissipation of the readout to $\approx
50$~mW/cm$^2$. Such low power consumption can be achieved by means of
power pulsing, i.e. turning off most components on the readout chips
during the 20~ms gaps between bunch trains.

Time slicing of hits with an accuracy of $\approx 10$~ns will be
required to separate beam-induced backgrounds from physics events.

\section{Detector concepts}\label{sec:det-concepts}
The two detector concepts CLIC\_ILD and CLIC\_SiD are currently under study.
They are adaptions of the two validated detector concepts  
ILD~\cite{ILD-LOI} and SiD~\cite{SiD-LOI}, developed for the
International Linear Collider (ILC) [5] with a centre-of-mass energy
of 500~GeV. 
The main CLIC-specific adaptions to the ILC detector concepts
are an increased hadron-calorimeter depth (7.5 $\Lambda_i$)
to improve the containment
of jets at the CLIC centre-of-mass energy of up to 3 TeV, and a
re-design of the vertex and forward regions to mitigate the effect of
high rates of beam-induced backgrounds.

Both
detectors have a barrel and endcap geometry with the barrel
calorimeters and tracking systems located inside a superconducting
solenoid providing an axial magnetic field of 4~T in case of CLIC\_ILD
and 5~T in case of CLIC\_SiD. In the
CLIC\_ILD concept, the tracking system is based on a large Time
Projection Chamber (TPC) with an outer radius of 1.8~m complemented by
an envelope of silicon strip detectors and by a silicon pixel vertex
detector. The all-silicon tracking and vertexing system in CLIC\_SiD is
more compact with an outer radius of 1.3~m.

\subsection{Vertex-detector layouts}
\begin{figure}[htbp]
 \centering
{\includegraphics[width=0.8\textwidth]
        {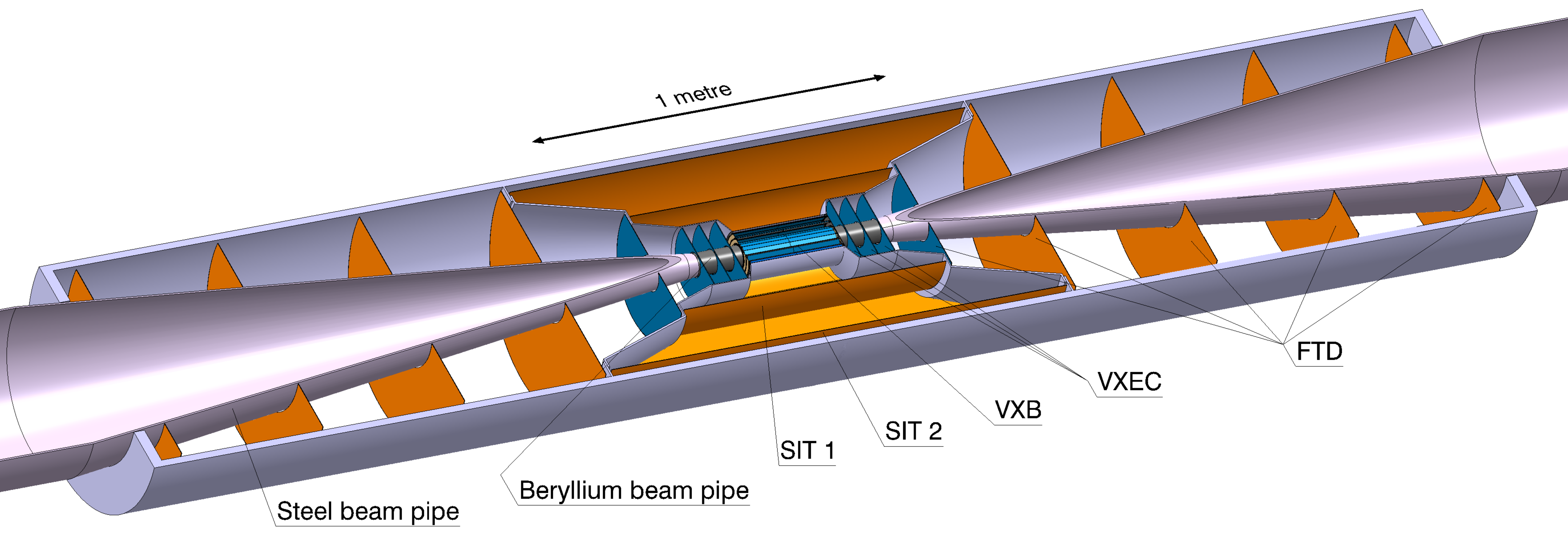}}
 \caption{View into the inner and forward tracking region of the
   CLIC\_ILD simulation model. Shown are the vertex barrel (VXB) and endcap (VXEC) pixel
layers, the two inner silicon barrel strip layers (SIT 1/2), the forward tracking disks (FTD),
the beam pipe, and the support shells for the silicon layers.}
\label{fig:vtx-layout-ild}
\end{figure}

Both CLIC detector concepts include 
silicon vertex detectors with $25 \mathrm{\mu}$m$\times 25\mathrm{\mu}$m pixel size.
They are integral parts of the full coverage tracking
systems, designed to improve the accuracy of the track reconstruction,
in particular for low transverse momenta.
In case of CLIC\_ILD, both the barrel and
forward vertex detectors consist of three double layers,  reducing
the material thickness needed for
supports. Figure~\ref{fig:vtx-layout-ild}
shows a sketch of
the inner tracking region of CLIC\_ILD.  For CLIC\_SiD, a geometry with
five single barrel layers and 7 single forward layers was chosen.

For CLIC\_ILD, the central beryllium beam-pipe is placed at a
radius of 29 ~mm, while the larger magnetic field in CLIC\_SiD leads to
a larger suppression of low-pT charged particles from beam-induced
background and therefore allows
for a reduced radius of the beam-pipe of 25~mm.
Stainless steel conical sections with a wall thickness of 4 mm extend
in the forward and backward directions and provide shielding against
backscattering backgrounds.

\subsection{Vertex-detector performance optimisation}
The vertex-detector performance with the baseline geometries has been
evaluated in Geant4-based \cite{geant4} full-detector simulation studies. 
The achieved impact-parameter resolutions 
are as precise as $3\mathrm{\mu}$m
for high-momentum tracks and the momentum resolution of the overall
tracking systems reaches the required value of $\sigma p_T/p_T^2\approx
2 \times 10^{-5}$GeV$^{-1}$.

\begin{figure}[htbp]
  \centering {
\includegraphics[width=.4\textwidth]
   {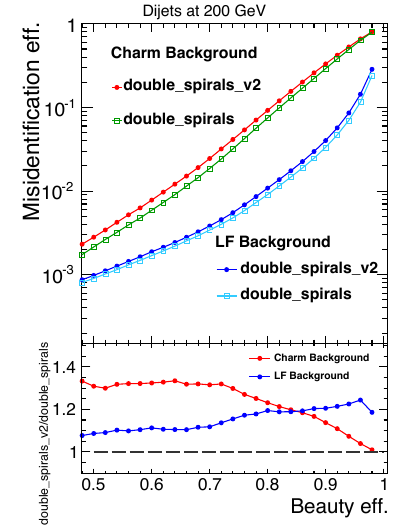}
\includegraphics[width=.4\textwidth]
   {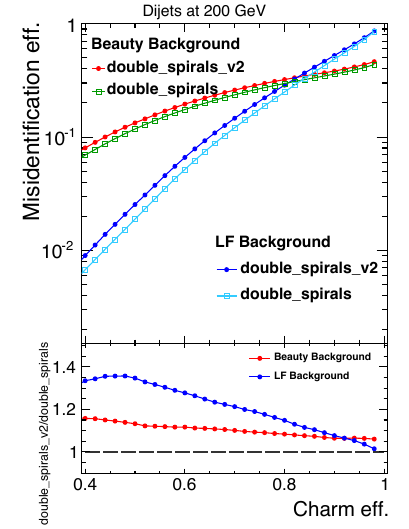}
}
  \caption{Flavor-tagging performance for two double-layer detector layouts with a 
spiraling endcap geometry and with different amounts of material 
(double\_spirals and double\_spirals\_v2). 
Shown is the background misidentification probability
(light flavors and c/b backgrounds) as a function of the signal efficiency for 
b-tagging (left) and c-tagging (right) of 200~GeV
dijet events. The bottom panels show the ratio of the misidentification probabilities 
(double\_spirals\_v2 / double\_spirals).}
  \label{fig:bc-tagging-material}
\end{figure}

Fast parametric simulation studies 
have been performed, to evaluate the effect of changes in the assumed
pixel size, material budget, arrangement of sensitive layers, and
inactive material~\cite{vtx-trk-optim,fwd-trk-optim}.  
In addition, Geant4-based full-detector simulations 
have been performed for selected geometries, probing the influence
of key design choices and detector parameters on the expected flavor-tagging 
performance~\cite{nilou-phd}.  The multi-variate flavor-tagging 
package  LCFIPlus~\cite{lcfiplus} is used
for this study, with dedicated training and testing samples for each geometry. 
Figure~\ref{fig:bc-tagging-material} shows the b-tagging and c-tagging
performance for 200~GeV dijet events for two double-layer detector layouts 
with a spiraling endcap geometry. 
The default layout with $\approx 0.2\%X_0$ per double layer 
(double\_spirals) is compared to a layout with an increased material budget of $\approx 0.4\%X_0$
per double layer (double\_spirals\_v2).
The fake rates increase by between approximately 5\% and 35\% for
the double\_spirals\_v2 layout, compared to the
default layout. 

The effect of changes of the flavor-tagging performance on the physics performance of the detectors was
estimated for the production of SM Higgs bosons at 3~TeV centre-of-mass energy in the process 
$e^+e^- \rightarrow H\nu\bar{\nu}$, with subsequent decay of the Higgs bosons in pairs of
b and c quarks~\cite{LCD-Note-2011-036}. The change in precision of the
$\sigma \times$BR measurement is estimated for a 20\% change in the fake rates of the dominating
light-flavor background, which results in a 6-7\% change in the
precision of the $H\rightarrow bb$ decay channel and in a 15\% change in the precision
of the $H\rightarrow cc$ decay channel.

\section{Hybrid readout technology}
The R\&D on pixel sensors and readout is focused on hybrid solutions,
combining high-resistivity sensors
with high-performance readout ASICs.
The target thickness for both the
sensor and readout layers is only 50~$\mathrm{\mu}$m each.  Slim-edge sensor
designs are under study and Through-Silicon Via (TSV) technology is foreseen for
vertical interconnection.  The hardware R\&D on
sensors and readout is complemented by silicon signal simulations, to
evaluate the impact of the technological parameters on the detector
performance under various operating conditions.

\subsection{Thin-sensor assemblies}
Planar pixel sensors with $55\mu$m pitch and different thicknesses (50-300 $\mu$m)
were procured from two different vendors.
Assemblies with Timepix readout ASICs (100 and 450 $\mu$m thickness)
were produced and characterised in laboratory measurements and beam tests. 
Slim-edge sensor designs (250-450 $\mu$m, two guard rings) are compared to designs 
with active edges (20-50 $\mu$m, one guard ring above the edge pixels).
Preliminary results show very good efficiencies in both cases, extending beyond the
edge pixels. Single-point resolutions of approximately 3~$\mu$m have been extracted
for clusters of two pixels using charge interpolation and taking into account non-linear charge sharing. 

Alternative sensor concepts possibly suitable for the CLIC vertex detectors 
include Low-Gain Avalanche Detectors (LGAD) with charge
multiplication~\cite{LGAD} and active sensors with capacitive charge coupling, 
implemented in a high-voltage CMOS process (HV-CMOS CCPD)~\cite{HVCMOS}.
%

\subsection{CLICpix readout chip}
The CLICpix hybrid readout chip~\cite{clicpix} will be implemented in a 65~nm CMOS process. 
The pixel size is $25\times25~\mathrm{\mu}$m. Simultaneous
4-bit Time-Of-Arrival (ToA) and Time-Over-Threshold (ToT) measurements
are implemented in each pixel, allowing for a front-end time slicing
with less than 10~ns and for analog readout to improve the position resolution. 
A photon counting mode allows for threshold equalization.
A compression logic is implemented with three selectable readout
modes: (1) no compression; (2) pixel-to-pixel
compression; (3) pixel-, cluster- and column-based compression.
The full chip can be read out in less than
800~$\mathrm{\mu}$s (for 10\% occupancy), using a 320~MHz readout clock. The
power consumption of the chip is dominated by the analog frontend
with a peak power corresponding to 2~W/cm$^2$. The total average power
consumption can be reduced to a value below the target of 50~mW/cm$^2$
by means of power gating for the analog part and clock gating for the digital part.

\begin{figure}[htbp]
 \centering
{\includegraphics[width=0.7\textwidth]{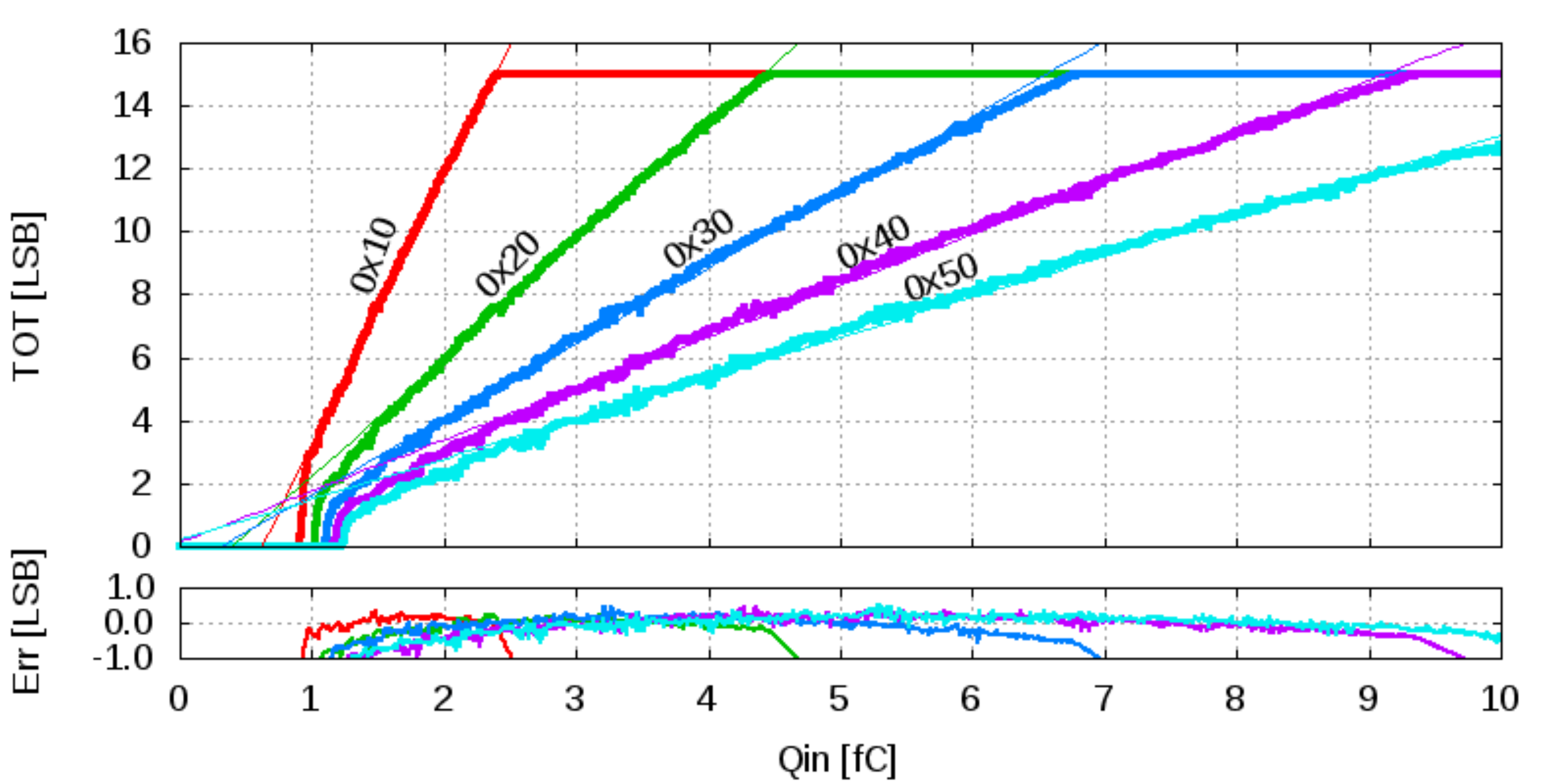}}
 \caption{ToT counts of one pixel, averaged over 256 samples,
 as a function of the injected charge and for different values for the feedback current. 
The bottom plot shows the deviation from a linear fit.}
\label{fig:itot_ikrum}
\end{figure}

A CLICpix demonstrator chip has been produced,
including a $64\times64$ pixel matrix. 
Readout tests have confirmed that 
the chip is fully functional and the performance is in agreement with 
simulations~\cite{clicpix-twepp-2013}.
Figure~\ref{fig:itot_ikrum} shows the result of a ToT scan for various values of the
feedback current. Higher feedback currents correspond to a faster return to the baseline,
thus increasing the dynamic range of the energy measurement.

Table~\ref{tab:clicpix-summary} summarizes the results of the characterization measurements
and compares them to the expectations from simulations. Good agreement between measurements
and simulations is observed. 
\begin{table}[tbp]
\caption{Comparison of simulated and measured parameters of the CLICpix demonstrator chip.}
\label{tab:clicpix-summary}
\smallskip
\centering
\begin{tabular}{|lcc|}
\hline
Parameter&Simulated Value&Measured Value\\
\hline
ToA Accuracy & $<10$~ns & $<10$~ns\\
Gain & $44$~mV/k$e^-$ & 40 mV/k$e^-$\\
Dynamic Range & up to 40 k$e^-$ & up to 45 $ke^-$\\
Equivalent Noise (bare chip) & $\sigma =60e^-$ & $\sigma =51 e^-$ (average)\\
DC Spread (uncalibrated) & $\sigma =160e^-$& $\sigma =128e^-$\\
DC Spread (calibrated) & $\sigma =24e^-$ & $\sigma =22e^-$\\
Minimum threshold & 388 $e^-$ & 417 $e^-$\\
Power consumption per pixel & 6.5 $\mathrm{\mu}$W & 7 $\mathrm{\mu}$W\\
\hline
\end{tabular}
\end{table}

\subsection{Through-Silicon Via (TSV) technology}
Through-Silicon Via (TSV) vertical interconnects remove the need for wire 
bonding connections on the side of the readout ASICs
and therefore allow for an efficient tiling to form larger modules with mimal 
inactive areas. 

A ``via last'' TSV process developed in collaboration with
CEA-LETI has demonstrated the feasibility of TSVs on functional
detector chips from the Medipix/Timepix chip family~\cite{TSV-leti}.
The project uses Medipix3 readout wafers produced in 130~nm CMOS
technology, which include landing pads for the I/O circuitry, as
well as dedicated TSV test structures at the periphery of the wafers.

The ``via last'' process proceeds in the following steps: (1)
deposition of the Under-Bond-Metalisation (UBM) on the front-side of
the chips; (2) temporary bonding of the front side to a support wafer
and thinning to 120~$\mu$m from the back side; (3) etching and isolation of vias with 
240~$\mu$m diameter; (4)
deposition of a 5~$\mu$m thick copper layer and patterning of the via trenches; 
(5) passivation and UBM on the backside; (6) de-bonding of the front-side support
wafer and attachment of the back side to dicing tape.

Tests on the processed wafers show good results, with a low resistivity of the
vias (<1~$\Omega$) and a sufficient isolation to the outside (leakage current
<1~$\mathrm{\mu}$A at 1~V). Functional tests of the chips before and after TSV
processing indicate no significant deterioration of the performance.
An ongoing second phase of the TSV project aims at demonstrating 
a good yield of the process.

\subsection{Simulation of signals in silicon}\label{sec:silicon-simulation}
Monte-Carlo
simulations and validations in test beams are performed to study the
signal development in the silicon sensors and extract parameters such
as the charge rise time, collection efficiency, charge sharing between
pixels and signal/noise ratio. The dependence on the type,
momentum and incident angle of the particles and on the electric and
magnetic field are studied using TCAD~\cite{TCAD} and 
Monte-Carlo charge-transport~\cite{charge-collection-simulation,MC-charge-transport}
simulations. 

The combination of the electric and magnetic field in the sensors
leads to a spread of the charge cloud due to the Lorentz-angle effect, 
thus affecting the charge sharing between neighboring pixels in the barrel
layers.
This effect will be of particular importance in the very thin sensors
foreseen for the CLIC vertex detector. For a p-in-n sensor with a
resistivity of 10~k$\Omega$cm and a thickness of 50~$\mathrm{\mu}$m, the
depletion voltage is only $V_{dep}\approx1$~V, leading to an average
field in the sensors of only 200~V/cm. The resulting calculated Lorentz angle
between the direction of the electric field lines and the direction of
the drifting charge carriers is 33$^\circ$ (7$^\circ$) for electrons
(holes), assuming a magnetic field of 4~T and a
temperature of 27~$^\circ$C~\cite{lorentz-calculation}.
For an increased operation voltage of 40~V, the Lorentz angle
is reduced to 19$^\circ$ (6$^\circ$) for electrons (holes). 
Such values have a seizable effect on the detector resolution
and therefore need to be taken into account in the choice of the
readout polarity (electrons or holes), the layer placement, and the
sensor operation voltage.

\section{Detector integration}\label{sec:det-integration}
The detector performance requirements lead to challenging constraints
for the mechanical and electrical integration of the vertex-detector
components. The powering, cooling, mechanical supports and
the service lines are therefore addressed in an integrated approach at an early
stage of the detector design.

\subsection{Powering}\label{sec:powering}
The ambitious power-consumption target of
less than 50~mW/cm$^2$ in the vertex detectors can only be
achieved by means of pulsed powering, taking advantage of the low duty
cycle of the CLIC machine. The main power consumers in the readout
circuits will be kept in standby mode during most of the gap of
20 ms between consecutive bunch trains. Furthermore, efficient power
distribution will be needed to limit the amount of material used for
cables. Both the
power pulsing and the power-delivery concepts have to be designed and
thoroughly tested for operation in a magnetic field of 4-5~T.

A powering scheme based on constant current sources,
Low Drop-Out regulators (LDOs) and local energy storage with silicon
capacitors has been proposed for the power
delivery and power pulsing of the CLIC vertex detectors~\cite{fuentes-twepp2013}.
The scheme takes advantage of the low duty cycle of the CLIC machine,
to limit the current and thereby
the cabling material needed to bring power to the detectors.

\begin{figure}[htbp]
  \centering {
\includegraphics[width=.8\textwidth]
   {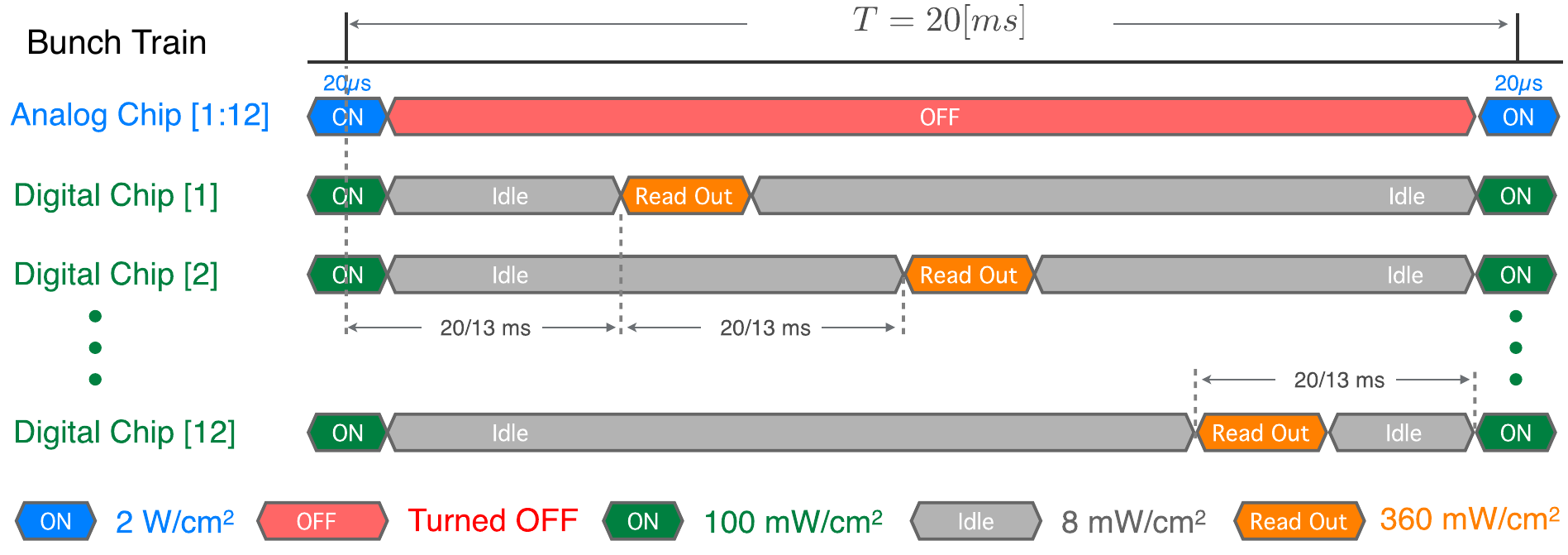}
}
  \caption{
    Analog and digital power consumption states during a CLIC bunch-train cycle for the 12 CLICpix chips in a 
half ladder. 
}
  \label{fig:an-dig-power}
\end{figure}

The ladders of the barrel vertex detector are assumed to consist of 24 readout
chips covering a surface area of approximately 24~cm$^2$. Powering and readout
will be connected on either side.
Figure~\ref{fig:an-dig-power} shows the
analog and digital power consumption states during a CLIC bunch-train cycle 
for the 12 CLICpix chips in a half ladder.
The nominal
operation voltage for both the analog and digital components in the
chips is 1.2~V.
The minimum time needed with stable conditions in the analog
components is approximately 20~$\mathrm{\mu}$s, in which both digital and
analog power are at their peak values of 48~W and 2.4~W,
respectively. The rise and fall times can be tuned by design
of the readout chips within certain limits and are both assumed to be of the order of
1~$\mathrm{\mu}$s.
During the remaining time of the 20~ms cycle the analog
power can be switched off. For the digital components, on the other
hand, an average continous power of 0.54~W per ladder is still required,
in order to sequentially send the data off detector and
keep the chips in a standby mode.

\begin{figure}[htbp]
 \centering
{\includegraphics[width=.99\textwidth] {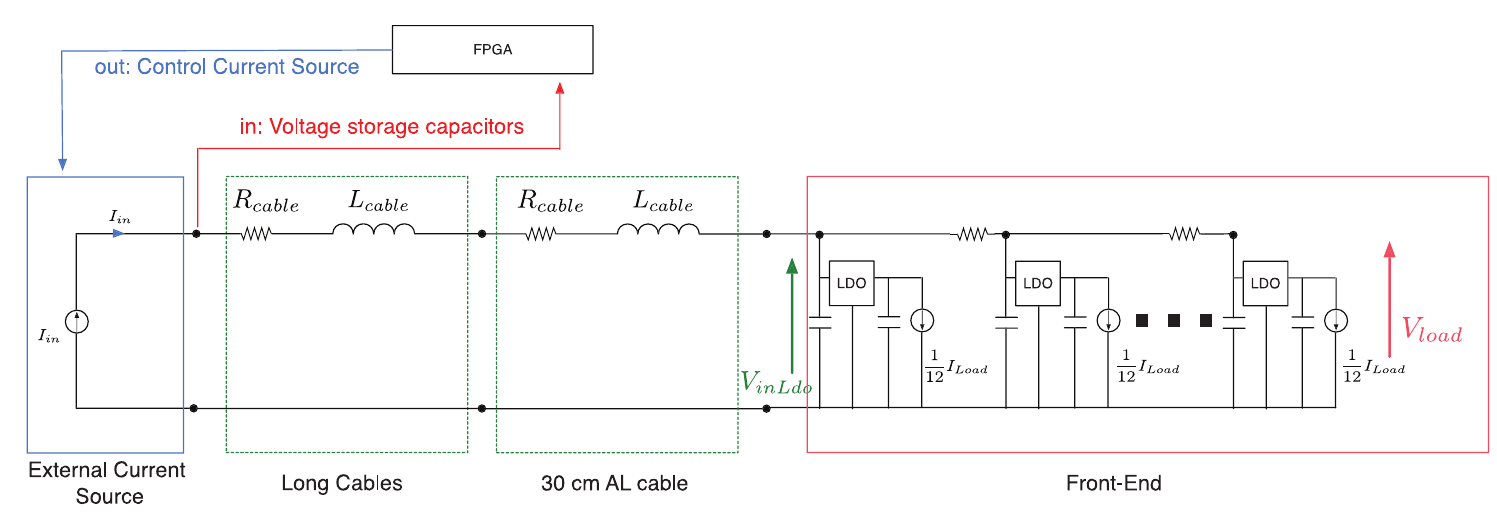}}
 \caption{Powering scheme for one half ladder in the barrel vertex
   detector, consisting of 12 CLICpix readout chips.}
\label{fig:powering-scheme}
\end{figure}

Figure~\ref{fig:powering-scheme} shows a sketch of the proposed
powering scheme for a half ladder in the
barrel vertex detector.
A programmable current source located in the back end provides a constant current 
to charge up the on-detector storage capacitors. 
The small duty cycle allows for a low continous current of only approximately 
20 mA for a half ladder.  
Flex cables
consisting of aluminum conductors on a Kapton substrate bring the
power to storage capacitors mounted on the individual CLICpix chips of the half
ladder. LDOs are used to provide a constant voltage
of 1.2~V (1~V) to the analog (digital) components of the chips. The LDOs for the analog part are
only operated during 250~$\mathrm{\mu}$s around the 20~$\mathrm{\mu}$s acquisition time,
while the LDOs for the digital part are always on.
The voltage at the LDO input is continously sensed at the back end and used as feedback for
an FPGA controlling the programmable current source. This way the load current is adapated 
after each cycle, to provide enough charge to the storage capacitors for the next
high-power period.

\begin{figure}[htbp]
 \centering
{\includegraphics[width=.99\textwidth] {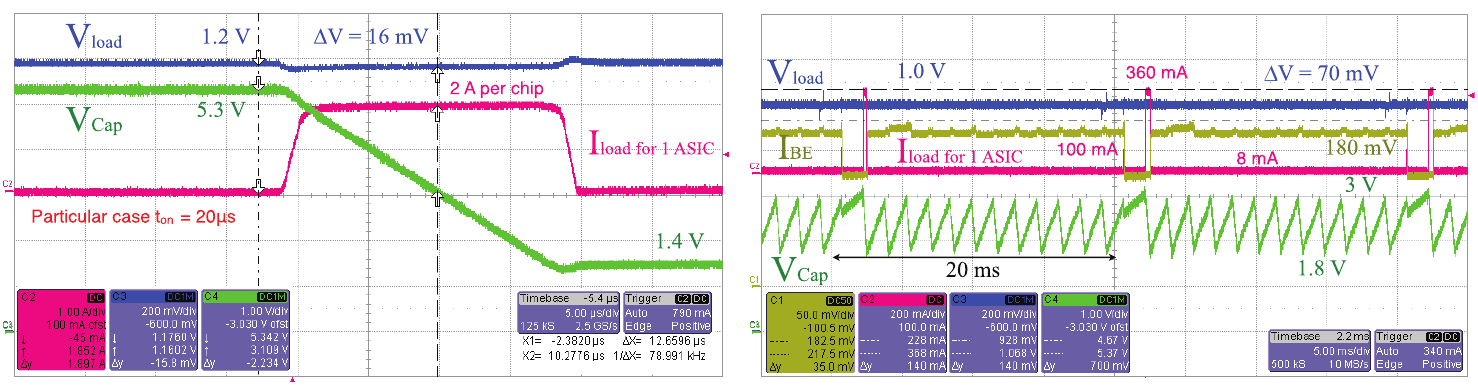}}
 \caption{Measurement results for the analog (left) and digital (right) components
in a half ladder.}
\label{fig:powerpulsing-measurements}
\end{figure}

A mockup has been built and operated to test the proposed powering scheme.
In this model the CLICpix chips are replaced with programmable loads emulating the 
current consumption of the chips. 
Figure~\ref{fig:powerpulsing-measurements} shows the achieved power regulation for both
the analog and digital components of a half ladder.
The voltage regulation for the analog components is stable at the level of approximately 16~mV,
while the less critical regulation of the digital voltage is approximately 70~mV.
With this scheme, the current into the half ladder is limited to less than 300~mA and a 
total power dissipation of approximately 45~mW/cm$^2$ is achieved.
The measurements are in agreement with equivalent-circuit simulations.

The average contribution of the flex cable, the LDOs and the silicon
capacitors to the material in the ladder area corresponds to 0.1\%$X_0$ per single
detection layer.
The use of thinner conductors and future improvements of the
silicon-capacitor technology are expected to decrease the material budget
even further. 

\subsection{Cooling}\label{sec:cooling}
A total power of approximately 500 W will be dissipated in the vertex
detectors alone. The small material budget for the inner 
tracking-detectors constrains severely the permitted amount of cooling
infrastructure. For all pixel layers, forced air-flow cooling
is therefore foreseen.

\begin{figure}[htbp]
 \centering
{\includegraphics[width=.35\textwidth]{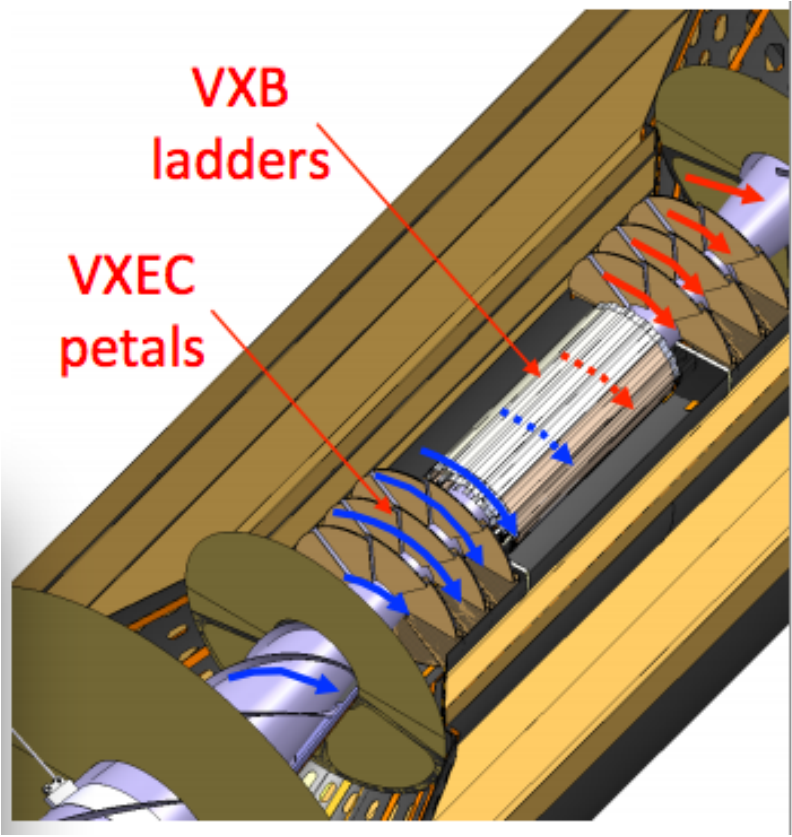}}
\hspace{0.5cm}
{\includegraphics[width=.5\textwidth]{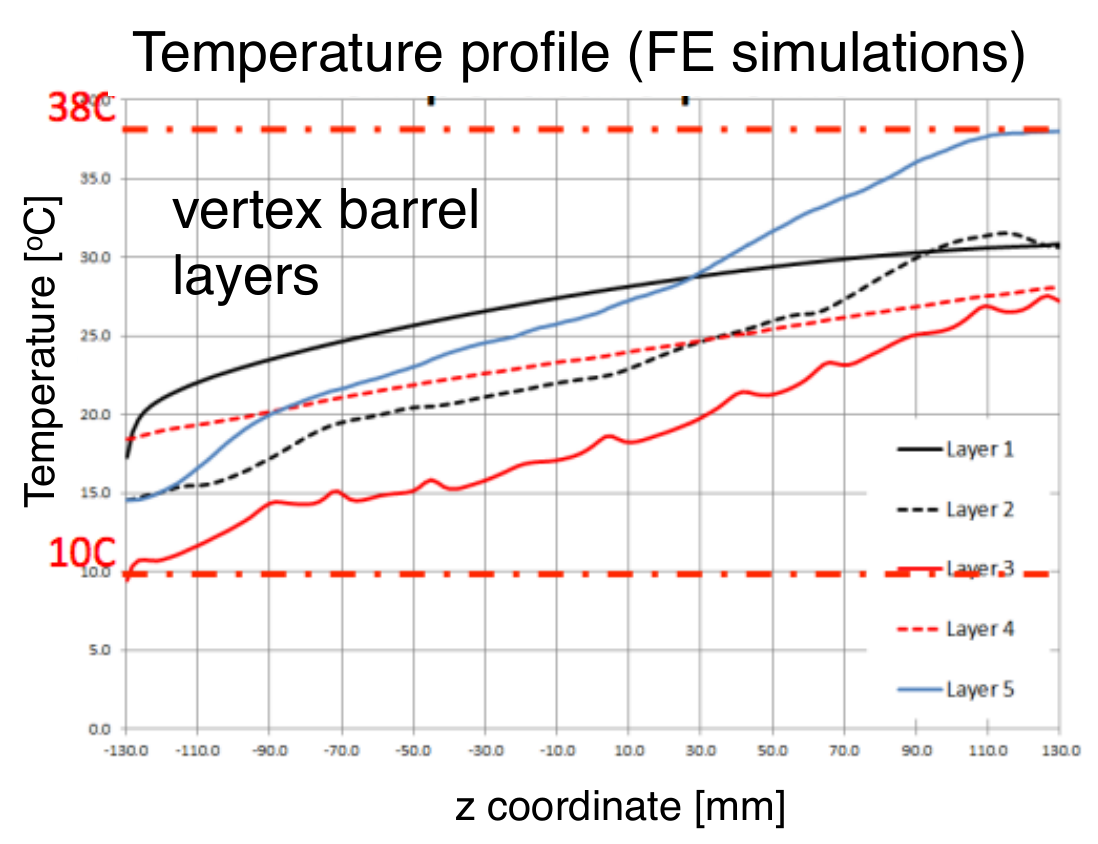}}
 \caption{Air flow in a vertex detector layout with spiral endcap geometry (left) 
and finite element simulation of the resulting temperature profile in the
barrel layers (right).} 
\label{fig:cooling}
\end{figure}

Feasibility studies have been performed for an inner-detector
cooling system with sufficient heat removal 
capability~\cite{LCD-Note-2013-007}.
A spiral arrangement of the endcap pixel disks allows
for air flow through the disks on one side, into the barrel region and
out through the endcap-disks on the other
side (Fig.~\ref{fig:cooling} (left)). Figure~\ref{fig:cooling} (right) shows the resulting
temperature profile in the pixel detector barrel layers, obtained from an
ANSYS finite element simulation. With an air temperature of
0~$^\circ$C and an average flow velocity of 11~m/s at the end-cap inlet, the
temperature in the innermost barrel layer reaches up to 40~$^\circ$C and
stays below 30~$^\circ$C in the other layers. The temperature span
between inlet and outlet is up to 20~$^\circ$C. Heat transfer
through conduction has not been taken into account in the simulation.

Further R\&D is ongoing to
demonstrate the feasibility of this air-flow cooling scheme using a thermo-mechanical
mockup. A simplified geometry consisting of a single 
stave mockup with heating elements has been implemented inside a wind tunnel.
Temperature and vibration amplitudes are monitored
as function of air-flow velocity and the results are compared to finite-element simulations.

\subsection{Mechanical integration}\label{sec:mechanical-integration}
Low-mass mechanical solutions are under study, to provide sufficient
support for the sensors, the readout chips and the cabling
infrastructure, while leaving enough clearance for air-flow cooling.
Realistic assembly and in-situ testing scenarios taking into account the 
constraints from the surrounding detector elements have been developed. 
Various prototype stave supports made of carbon-fibre reinforced polymers (CFRP) 
have been produced and tested for bending stiffness. The results of the tests are
compared to analytical calculations and to finite-element simulations. 
Support structures including silicon carbide (SiC) foam are also under study. 

\section{Conclusions}
The CLIC machine environment and the requirements for precision
physics measurements place challenging demands on the vertex-detector
systems. Initial detector layouts meeting these demands have been proposed
and are currently being refined in line with results from detector-optimisation and hardware 
development studies.
An active R\&D program on sensor and readout technologies,
simulation, power delivery and power pulsing, mechanical integration and
cooling is in place.

%
%




\begin{thebibliography}{99}
 \bibitem{CDRVol1} \emph{A Multi-TeV linear collider based on CLIC technology: CLIC Conceptual Design Report},
edited by M. Aicheler, P. Burrows, M. Draper, T. Garvey, P. Lebrun, K. Peach, N. Phinney,
H. Schmickler, D. Schulte and N. Toge, 2012, \href{https://cds.cern.ch/record/1500095}{CERN-2012-007}.
 \bibitem{CDRVol2} \emph{Physics and Detectors at CLIC: CLIC Conceptual
   Design Report}, edited by L. Linssen, A. Miyamoto, M. Stanitzki,
   H. Weerts, 2013, \href{https://cds.cern.ch/record/1425915}{CERN-2012-003}.
 \bibitem{CDRVol3} \emph{The CLIC Programme: towards a staged $e^+e^-$
   Linear Collider exploring the Terascale, CLIC Conceptual Design
   Report}, edited by P. Lebrun, L. Linssen, A. Lucaci-Timoce,
 D. Schulte, F. Simon, S. Stapnes, N. Toge, H. Weerts, J. Wells,
2012,
 \href{https://cds.cern.ch/record/1475225}{CERN-2012-005}.
\bibitem{background-note} D. Dannheim and A. Sailer,
  \emph{Beam-Induced Backgrounds in the CLIC Detectors}, 2011,
 \href{http://cdsweb.cern.ch/record/1443516}{LCD-Note-2011-021}.
\bibitem{ILD-LOI} T.~Abe et al., \emph{The International Large Detector:
  Letter of Intent}, 2010, \href{http://arxiv.org/abs/1006.3396}{arXiv:1006.3396}.
\bibitem{SiD-LOI} H. Aihara et al., \emph{SiD Letter of Intent}, 2009,
  \href{http://arxiv.org/abs/0911.0006}{arXiv:0911.0006}, SLAC-R-944.

\bibitem{geant4} S. Agostinelli et al., \emph{Geant4 - a simulation toolkit}, NIM {\bf A 506} (2003) 250.

\bibitem{vtx-trk-optim}
D. Dannheim and M. Vos, \emph{Simulation studies for the layout of the vertex and tracking regions of the CLIC detectors},
2011, \href{https://cds.cern.ch/record/1443503}{LCD-Note-2011-031}.

\bibitem{fwd-trk-optim}
S.~Aplin et al.,
\emph{Forward tracking at the next $e^+e^-$ collider, Part II: experimental challenges and detector design},
JINST {\bf 8} (2013) T06001.

\bibitem{nilou-phd} N. Alipour Tehrani, \emph{Performance-Optimization Studies for the CLIC Vertex Detector},
2013, \href{https://cds.cern.ch/record/1606436}{CERN-THESIS-2013-149}.

\bibitem{lcfiplus}
T. Tanabe and T. Suehara, \emph{LCFIPlus}, 2013,
\href{https://confluence.slac.stanford.edu/display/ilc/LCFIPlus}{https://confluence.slac.stanford.edu/display/ilc/LCFIPlus}.

\bibitem{LCD-Note-2011-036}
T. Lastovicka, \emph{Light Higgs Production and Decays to Pairs of Bottom and Charm Quarks at 3 TeV},
2011,
\href{https://cds.cern.ch/record/1499128}{LCD-Note-2011-036}.

\bibitem{LGAD} N. Cartiglia et al., \emph{Performance of Ultra-Fast Silicon Detectors}, \href{http://arxiv.org/abs/1312.1080}{arXiv:1312.1080 }.

\bibitem{HVCMOS}  I. Peric et al.,
  \emph{High-voltage pixel detectors in commercial CMOS technologies for ATLAS, CLIC and Mu3e experiments},
 NIM~{\bf A731} (2013) 131.

\bibitem{clicpix}
P. Valerio, R. Ballabriga and M. Campbell,
\emph{Design of the 65 nm CLICpix demonstrator chip},
\href{https://cds.cern.ch/record/1507691}{LCD-note-2012-018}.
\bibitem{clicpix-twepp-2013} 
P. Valerio et al.,
\emph{A prototype hybrid pixel detector ASIC for the CLIC experiment},
 \href{https://cds.cern.ch/record/1623863}{CLICdp-Conf-2013-003}.

\bibitem{TSV-leti} D. Henry et al., \emph{TSV last for hybrid pixel detectors: Application to particle physics and imaging experiments}, Electronic Components and Technology Conference (ECTC), 2013 IEEE 63rd, pp.568, 28--31 May 2013, 
\href{http://dx.doi.org/10.1109/ECTC.2013.6575630}{doi: 10.1109/ECTC.2013.6575630}.

\bibitem{TCAD} Technology Computer Aided Design (TCAD),
  \href{http://www.synopsys.com/Tools/TCAD/}{http://www.synopsys.com/Tools/TCAD/}.

\bibitem{charge-collection-simulation} M. Benoit and L.A. Hamel,
  \emph{Simulation of charge collection processes in semiconductor CdZnTe $\gamma$-ray detectors}, 
NIM  {\bf A 606} (2009) 508.

\bibitem{MC-charge-transport} T.~Janssen, \emph{Simulation of charge
    collection processes in semiconductor silicon tracking detectors},
 \href{https://edms.cern.ch/document/1240445/}{LCD-Open-2012-010}.
%
\bibitem{lorentz-calculation} V. Bartsch et al., \emph{An algorithm
    for calculating the Lorentz angle in silicon detectors},  NIM {\bf A 497} (2003) 389.

\bibitem{fuentes-twepp2013} 
G. Blanchot, D. Dannheim and C. Fuentes, \emph{Power pulsing schemes for vertex detectors at CLIC},
 \href{https://cds.cern.ch/record/1624088}{CLICdp-Conf-2013-005}.

\bibitem{LCD-Note-2013-007}
F. Duarte Ramos, H. Gerwig, M. Villarejo Bermudez,
\emph{CLIC inner detectors cooling simulations},
2013,
\href{https://cds.cern.ch/record/1572989}{LCD-Note-2013-007}.


%
%
%
%
%

\end{thebibliography}
\end{document}